\documentclass[conference]{IEEEtran}
\usepackage{cite}
\usepackage[cmex10]{amsmath}
\usepackage{array}
\usepackage{mdwmath}
\usepackage{mdwtab}
\usepackage{fixltx2e}
\usepackage{url}
\usepackage{psfrag}
\usepackage{graphicx}
\usepackage{xcolor}
\usepackage{flushend}
\usepackage{float}
\usepackage[caption=false,font=footnotesize]{subfig}
\usepackage{multirow}
\newcommand{\nfar}{\mathrm{F}}
\newcommand{\nnear}{\mathrm{N}}
\newcommand{\nap}{\mathrm{A}}

\newcommand{\snfar}{\mathrm{f}}
\newcommand{\snnear}{\mathrm{n}}

\newcommand{\srelay}{\mathrm{r}}

\newcommand{\betam}{\overline{\beta}}
\newcommand{\power}{\mathcal{P}}
\newcommand{\ttime}{\mathcal{T}}
\newcommand{\stime}{\mathcal{S}}

\newcommand{\tsucc}{t_{\mathrm{sc}}}
\newcommand{\tsuccf}{t_{\mathrm{sc,f}}}
\newcommand{\tsuccn}{t_{\mathrm{sc,n}}}
\newcommand{\tsuccr}{t_{\mathrm{sc,r}}}

\newcommand{\fpacket}{\mathtt{t}_{\mathrm{f}}}
\newcommand{\npacket}{\mathtt{t}_{\mathrm{n}}}
\newcommand{\rpacket}{\mathtt{t}_{\mathrm{r}}}
\newcommand{\cpacket}{\mathtt{t}_{\mathrm{c}}}

\newcommand{\ttransf}{t_{\mathrm{f}}}
\newcommand{\ttransn}{t_{\mathrm{n}}}
\newcommand{\ttransr}{t_{\mathrm{r}}}

\newcommand{\tcoll}{t_{\mathrm{c}}}
\newcommand{\tidle}{t_{\mathrm{i}}}

\newcommand{\psucc}{p_s}
\newcommand{\pcoll}{p_\mathrm{c}}
\newcommand{\pidle}{p_\mathrm{i}}

\newcommand{\capacity}{\mathsf{C}}
\newcommand{\cdirect}{\capacity_{\mathrm{dir}}}
\newcommand{\ctwohop}{\capacity_{\mathrm{2h}}}
\newcommand{\cdf}{\capacity_{\mathrm{df}}}

\newcommand{\lefto}{\mathopen{}\left}

\newtheorem{remark}{Remark}

\IEEEoverridecommandlockouts

\begin{document}
\title{\ 
\\[-0.0in]
\frenchspacing Cooperative Protocols for Random Access Networks}
%\author{\IEEEauthorblockN{Georg B\"ocherer, Alexandre de Baynast}

\author{\IEEEauthorblockN{Georg B\"ocherer\IEEEauthorrefmark{1},
Alexandre de Baynast\IEEEauthorrefmark{2} and
Rudolf Mathar\IEEEauthorrefmark{1}}
\IEEEauthorblockA{\IEEEauthorrefmark{1}Institute for Theoretical Information
Technology\\
RWTH Aachen University,
52056 Aachen, Germany\\ Email: \{boecherer,mathar\}@ti.rwth-aachen.de}
\IEEEauthorblockA{\IEEEauthorrefmark{2}European Microsoft Innovation Center
(EMIC) GmbH\\
Ritterstrasse 23, 52072 Aachen, Germany\\
Email: alexdeba@microsoft.com}
\thanks{This work has been supported by the UMIC Research Centre, RWTH
Aachen University}
}

\maketitle
\begin{abstract}
Cooperative communications have emerged as a significant concept
to improve reliability and throughput in wireless systems.
On the other hand, WLANs based on random access mechanism
have become popular due to ease
of deployment and low cost. 
Since cooperation introduces extra transmissions among the cooperating nodes
and therefore increases the number of packet collisions, it is not clear whether
there is any benefit from using physical layer cooperation under random access.
In this paper, we develop new low complexity cooperative protocols for random access that outperform
the conventional non cooperative scheme for a large range of signal-to-noise ratios.
\end{abstract}

\section{Introduction}

Cooperative communications have emerged as a significant concept
to improve reliability and throughput in wireless systems~\cite{Cover1979,Sendonaris2003,Laneman2004,Kramer2005,Gupta2003}.
In cooperative communications, the
resources of distributed nodes are effectively pooled for the collective benefit of all nodes.
The broadcast nature of the wireless medium is the key property that allows for
cooperation among the nodes:~transmitted signals can, in principle,
be received and processed by any number of nodes.
Although these extra observations of the transmitted signals are available
for free (except, possibly, for the cost of additional energy consumption for sensing operation),
wireless network protocols often ignore or discard them.
The main reason for this is that additional transmissions among the cooperating nodes
are needed in order to efficiently pool their resources.
In large random access networks without centralized scheduler like in IEEE 802.11 DCF systems~\cite{Bianchi2000},
these extra transmissions will increase the number of packet collisions and it is not
clear whether there is any benefit of using physical layer cooperation in this case.
In the case of random access, cooperative strategies,
if handled poorly, can even cause performance degradation and a non cooperative
scheme,
which consists in transmitting the messages of all nodes
directly to the access point, might be preferrable.

In this paper, we take the first steps in understanding
the issues in designing practical cooperative communication systems for random access networks.
Specifically, we closely model the interaction
between the physical and medium access channel (MAC) layers
in case of physical layer cooperation by a finite state machine.
Our model is quite generic since it includes any cooperative or non cooperative multihop transmission scheme.
Based on this model, we develop and analyze three new protocols that take full advantage
of the node cooperation at the physical layer. We focus on
Decode-and-Forward protocols where the intermediate node $\nnear$ decodes the
full
message sent by the source and forwards only the information missing from the
original transmission needed by the
destination (here, the access point) to decode the original packet.
The Decode-and-Forward protocol was shown to considerably increase the
throughput \cite{Host-Madsen2005}.

The remainder of this paper is organized as follows. In
Section~\ref{sec:model}, we model the medium access channel
by taking the specifications of the physical layer cooperation into account.
In Section~\ref{sec:cooperative-protocols}, we develop two new simple cooperative protocols
that outperform the conventional approach.
The throughput analysis for these protocols is elaborated in Section~\ref{sec:throughput} and
performance results are discussed in Section~\ref{sec:discussion}.
Concluding remarks are presented in Section~\ref{sec:conclusions}.

\section{System Model}\label{sec:model}

We consider the network topology shown in Fig.~\ref{fig:terminals} where nodes
$\nfar$ and $\nnear$ send data
to the access point $\nap$, and in doing so, both nodes are
susceptible to mutually help each other.
In this study, we consider half-duplex relay channels~\cite{Host-Madsen2005},
i.e., the nodes
cannot transmit and receive simultaneously.

\subsection{Medium Access}
Throughout the paper, the nodes $\nfar$ and $\nnear$ transmit their messages to
node $\nap$ 
using the distributed coordination function (DCF) mechanism as in IEEE 802.11
standard \cite{Bianchi2000}.
In principle, other random access schemes such as Slotted Aloha
\cite{Bertsekas1992} can be analyzed in a similar way.
Under this assumption, no packet/sample synchronization between the nodes is
expected,
which greatly simplifies the implementation of the communication protocols.
Collisions may occur between $\nfar$ and $\nnear$ at the access point.
In order to avoid collisions, DCF adopts an exponential backoff scheme
with a discrete time backoff scale, in which a contention window
initiated with a minimum size can be adapted exponentially up to a
maximum size in case of collision. The length of a discrete timeslot
depends on the PHY specifications, a typical value being
$50\mu$s~\cite{Bianchi2000}.

\begin{figure}
\centering
\footnotesize
\psfrag{S}{$\nfar$}
\psfrag{R}{$\nnear$}
\psfrag{D}{$\nap$}
\psfrag{dSR}{$(1-\beta)$}
\psfrag{dRD}{$\beta$}
\psfrag{dSD}{$1$}
\includegraphics[width=0.8\columnwidth]{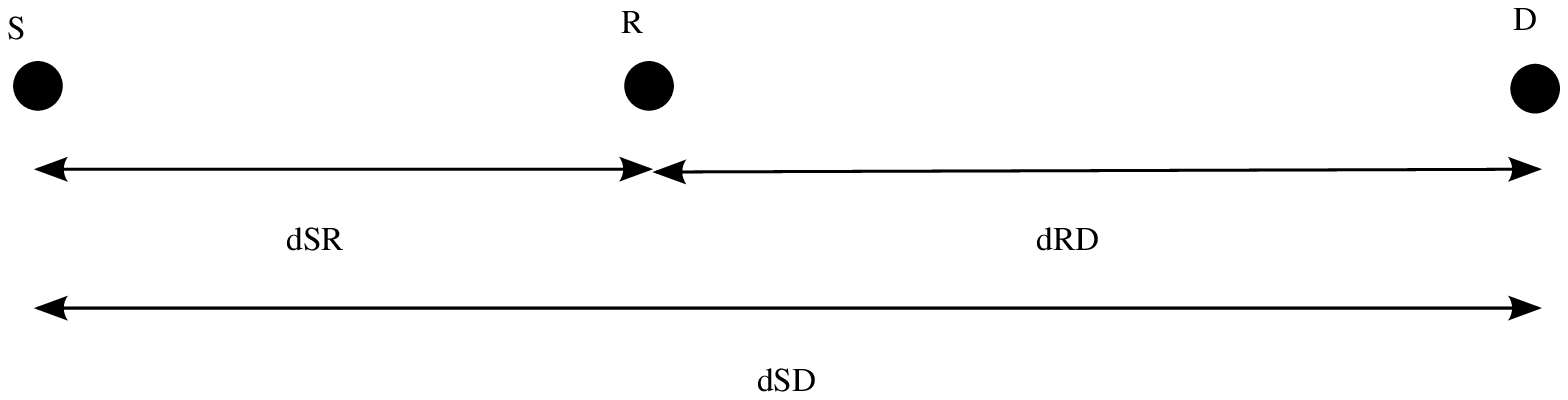}
\caption{The $3$-node relay channel. Node $\nnear$ serves as a relay for node
$\nfar$ as in \cite{Cover1979}. However, we assume here that the relay node
$\nnear$ additionally has its own data to transmit. For sake of simplicity,
$\nfar$, $\nnear$ and $\nap$ are assumed to be aligned. The distance between
$\nfar$ and $\nap$ is normalized to the unit. Nodes $\nnear$ and $\nap$ are
separated by distance
$\beta$.}
\label{fig:terminals}
\end{figure}

In the model shown in Fig.~\ref{fig:terminals}, we assume that the nodes operate
in saturation
conditions, i.e., they are backlogged and we do not need to consider packet
arrival
processes in our derivations.
Since all three nodes share the same wireless channel, the state of the network
can be described by the current channel state. We distinguish between three
phases:~first,
when node $\nfar$ or node $\nnear$ \textit{successfully transmits} a packet;
second, when a \textit{collision} between $\nfar$ and $\nnear$ occurs, and
third, when the channel is
\textit{idle}. Note that different phases can have different durations.
There are three types of transmission: $\nfar$
transmitting its own packet during the amount of time $\ttransf$, $\nnear$
transmitting
its own packet during $\ttransn$, and $\nnear$ relaying a packet
from $\nfar$ during $\ttransr$.
In our notations, the subscript $_\mathrm{sc}$ indicates that a transmission was
successful.
Similarly, we denote $\tcoll$ as the amount of time collisions occur and
$\tidle$ as the amount of time the channel is in idle state.
The duration $t$ of the observation time interval can thus be expressed as
\begin{align}
t=\tsucc+\tcoll+\tidle=\tsuccf+\tsuccn+\tsuccr+\tcoll+\tidle.
\label{eq:phaseDuration}
\end{align}
By normalizing the duration of each phase by the observation time interval $t$,
we can
express the normalized time division parameters as follows
\begin{align}
&\stime_\snfar=\frac{\tsuccf}{t},
\quad
\stime_\snnear=\frac{\tsuccn}{t},
\quad
\stime_\srelay=\frac{\tsuccr}{t},\nonumber\\
&\ttime_i=\frac{\tidle}{t},\quad
\ttime_c=\frac{\tcoll}{t},\quad
\ttime_\nfar=\frac{\ttransf}{t},
\quad
\ttime_\nnear=\frac{\ttransn}{t}.
\label{eq:throughputParameter}
\end{align}
The fractions of time
$\ttime_\nfar,\ttime_\nnear$ refer to the time $\nfar$ respectively $\nnear$
is transmitting, $\nfar$ is successfully transmitting during $\stime_\snfar$
and $\nnear$ is successfully transmitting its own packets during
$\stime_\snnear$ and successfully relaying during $\stime_\srelay$.
Clearly, $\stime_\snfar \leq \ttime_\nfar$, $\stime_\snnear \leq \ttime_\nnear$,
and $\stime_\srelay \leq \ttime_\nnear$ due to the collisions.
For sake of simplicity, we assume
that $\nfar$ and $\nnear$ are either idle or transmit with constant power,
e.g., $\nfar$ transmits either with power zero or with power
$\power/\ttime_\nfar$.
It can easily be verified
that~$\ttime_i+\ttime_c+\stime_\snfar+\stime_\snnear+\stime_\srelay=1$.

\subsection{Physical layer considerations}
\label{sec:model-PHY}

Under the above orthogonality between the channel states, we can now
conveniently, and without loss of generality, characterize our
channel models using a time-division notation.
We assume free-space path loss, i.e., the power of the propagating signal is
attenuated
with the source-destination distance to the power of $\gamma$. The coefficient
$\gamma$ denotes the pathloss exponent \cite[Chap. 2]{Tse2005}
with a typical range of $1.5 \leq \gamma \leq 4$. We utilize a
baseband-equivalent, discrete-time channel
model for the continuous-time channel. The distance between $\nfar$ and $\nap$
is
normalized to the unit. Denote $\betam = 1-\beta$ as the distance between nodes
$\nnear$ and $\nap$.
When $\nfar$ is transmitting (under our assumptions, meanwhile $\nnear$ and
$\nap$ are listening),
\begin{align}
y_\nnear[k]	&
=\betam^{-\gamma/2}x_\nfar[k]+z_\nnear[k]\label{eq:channelModelFirst} \\
y_\nap[k]	&	=x_\nfar[k]+z_\nap[k], \label{eq:y_nap}
\end{align}
where $x_\nfar$ is the signal transmitted by node $\nfar$.
The sequences $y_\nnear$ and $y_\nap$ represent the signals received at
node $\nnear$ and $\nap$, respectively.
The signals $z_\nnear$ and $z_\nap$ capture the effects of receiver noise and
other forms of interference in the system. We model them as
zero-mean mutually independent, circular symmetric, complex
Gaussian random sequences with variance $1$.
When $\nnear$ is transmitting and $\nap$ is listening, we model the channel as
\begin{align}
y_\nap[k]&=\beta^{-\gamma/2}x_\nnear[k]+z_\nap[k].\label{eq:channelModelLast}
\end{align}
\label{eq:model}
During the remaining time, both nodes $\nfar$ and $\nnear$ can simultaneously
transmit (collision) or remain idle.
In the case of collision, we assume that the access point cannot detect none of
the messages
and discards the received signal. Therefore, there is no need to model the
channel in this case.

Assuming that the transmitted signals $x_\nfar$ and $x_\nnear$ are subject to
the average
power constraints
\begin{align}
&\lim_{m\rightarrow\infty}\frac{1}{2m+1}\sum\limits_{k=-m}^m
|x_\nfar[k]|^2 \leq \power,\nonumber\\
&\lim_{m\rightarrow\infty}\frac{1}{2m+1}\sum\limits_{k=-m}^m
|x_\nnear[k]|^2 \leq \power,
\label{eq:average_power_constraint}
\end{align}
we parameterize the channel model by the signal-to-noise-ratios
$\power/(1-\beta)^{\gamma}$ between
$\nfar$ and $\nnear$, $\power/\beta^{\gamma}$ between $\nnear$ and $\nap$ and
$\power$ between $\nfar$ and $\nap$.

\section{Cooperative Protocols}
\label{sec:cooperative-protocols}

In this section, we describe three low-complexity cooperative
protocols that can be utilized in the network of Fig.~\ref{fig:terminals}.
All three protocols are subject to the same power
constraint~(\ref{eq:average_power_constraint}).

In our study, we are interested in protocols that optimize resource allocation
such that the flow with lowest rate is maximized.
We define the achievable minimum rate $C$ as the minimum rate granted over all flows.
In the transmission model in Fig.~\ref{fig:terminals},
there are two flows, one initiated by node $\nfar$ and one initiated by node
$\nnear$.
The maximum achievable minimum rate is determined by the flow with lowest rate:
\begin{equation}
C = \max_{T} \min\left\{ C_{\nfar} , C_{\nnear} \right\},
\label{eq:achievable-minimum-rate}
\end{equation}
where the maximum is taken over all possible time division configurations of the
network parameterized by the set
\begin{align}
T=(\ttime_\nfar,\ttime_\nnear,\stime_\snfar,\stime_\snnear,\stime_\srelay).
\end{align}

\subsection{Benchmark for cooperative schemes}

In order to evaluate the benefit of cooperation among the nodes $\nfar$ and $\nnear$,
we first determine the maximum achievable minimum rate for non cooperative schemes.
We consider two basic non cooperative schemes: the Direct-Link and the Two-Hop
schemes.

\subsubsection{Direct-Link}
The Direct-Link scheme has been successfully adopted by the standard IEEE
802.11, in which
each node communicates directly with the access point.
The max-min capacity~(\ref{eq:achievable-minimum-rate}) for the Direct-Link
transmission scheme is readily given by the
capacity formula for the additive white Gaussian noise (AWGN) channel~\cite{Cover2006}
with the corresponding SNR values for $\nfar$ and $\nnear$ as stated in Section~\ref{sec:model}:
\begin{align}
\cdirect=\max_T&\min\left\{
\stime_\snnear\log\lefto(1+\frac{\power}{\beta^\gamma\ttime_\nnear}\right),
\stime_\snfar\log\lefto(1+\frac{\power}{\ttime_\nfar}\right)
\right\}, \label{eq:capacityDirectLast}
\end{align}
where the first and second terms correspond to the achievable rate for node $\nnear$ and node $\nfar$, respectively.
Since both $\nfar$ and $\nnear$ are transmitting their data directly
to $\nap$, no relaying is needed and we have $\stime_\srelay=0$.

When node $\nfar$ is very far from the access point $\nap$, the rate of the link
between nodes
$\nfar$ and $\nap$ becomes the bottleneck of the achievable minimum rate. In
this case, it might be preferable
to consider the Two-Hop solution, which consists of first transmitting the
message from $\nfar$
to $\nnear$ and second forwarding it from $\nnear$ to $\nap$. 

\subsubsection{Two-Hop}
By applying the capacity formula for AWGN channels with the corresponding SNR values,
the achievable rate for the Two-Hop scheme can be expressed as:
\begin{align}
\ctwohop=&\max_T\min\left\{
\stime_\snnear\log\lefto(1+\frac{\power}{\beta^\gamma\ttime_\nnear}\right),
\stime_\snfar\log\lefto(1+\frac{\power}{\betam^\gamma\ttime_\nfar}
\right)\right.,\nonumber\\
&\qquad\left.\stime_\srelay\log\lefto(1+\frac{\power}{\beta^\gamma\ttime_\nnear}
\right)
\right\}
\label{eq:capacityTwohop}
\end{align}
where the first and second terms correspond to the achievable rate for the transmission
of the own data of nodes $\nnear$ and $\nfar$ to their respective one-hop neighbors $\nap$ and $\nnear$.
The last term represents the achievable rate for the flow of node $\nfar$ forwarded by $\nnear$.

\begin{remark}[MAC Considerations for the Two-Hop scheme]
The main challenge of designing a MAC protocol for the Two-Hop scheme resides in
the
\textit{coordination} strategy 
for $\nfar$ and $\nnear$. In order to complete the transmission initiated by $\nfar$,
$\nnear$ needs to forward the received packet to~$\nap$.
We propose here a very simple policy as follows.
Nodes $\nfar$ and $\nnear$ initially compete for the channel.
If $\nnear$ gains the channel access, it transmits its packet.
Once the transmission has been acknowledged by the access point, 
both nodes $\nfar$ and $\nnear$ compete again for the channel.
If $\nfar$ gains channel access, it transmits its packet to $\nnear$.
Under our policy, the node $\nnear$ is \textit{obliged} to tentatively decode the packet
and, if it succeeds, to put it first in its packet queue.
Next time $\nnear$ gains the channel access, it forwards the packet to $\nap$.
In order to keep $\nfar$ from flooding $\nnear$ with packets, $\nnear$ keeps
only one packet from
$\nfar$ at a time  (in first position of its queue). Consequently,
if $\nfar$ gains channel access and transmits its packet whereas $\nnear$ has still a packet to forward,
$\nnear$ will ignore the transmission of $\nfar$. This principle is illustrated
in Fig.~\ref{fig:states}(b).\\
\label{re:remark-queue}
\end{remark}

Assuming that the optimal decision of selecting Direct-Link or
Two-Hop scheme is taken by a routing protocol (AODV for instance),
the maximum achievable minimum rate for the non-cooperative case can be expressed as
\begin{equation}
C_{\mbox{\footnotesize no coop}} = \max\left\{ \cdirect, \ctwohop \right\}.
\label{eq:C_nocoop}
\end{equation}
In the sequel, the performance gain of cooperative protocols will be evaluated against~(\ref{eq:C_nocoop}).
The main idea behind the three cooperative protocols is to consider the Two-Hop
scheme without discarding the signal that has been sent by
$\nfar$ at the access point~(\ref{eq:y_nap}).

\subsection{Naive Decode-and-Forward protocol}

We first consider the basic Decode-and-Forward scheme in which both nodes
$\nfar$ and $\nnear$
have to send their own data to the access point $\nap$.
For sake of clarity, we first expose the strategy from the physical layer point of view.
We can distinguish the three phases in Table~\ref{tab:DF}.
In Phase~1, $\nnear$ directly sends its message to node $\nap$. In this phase, $\nfar$ cannot help.
In Phase~2, $\nfar$ sends its message to the intermediate node $\nnear$ such that $\nnear$ can decode the message.
Node $\nap$ receives the message but cannot decode it due to the larger distance
between $\nfar$ and $\nnear$.
However, contrary to the Two-Hop scheme, $\nap$ stores the received signal for
the next phase.
In Phase~3, $\nnear$ transmits only the missing information to $\nap$ such that
together with
the message previously received in Phase~2, $\nap$ can completely decode the message from $\nfar$.
During Phase~3, we assume that $\nfar$ remains idle for two reasons: first, the
throughput gain
by allowing $\nfar$ to transmit together with $\nnear$ is rather little especially if its distance to $\nap$ is large;
second, simultaneous transmissions of $\nfar$ and $\nnear$ require time synchronization at the
sample level, which is costly in practice.
Comparison of the different phases of the Decode-and-Forward protocol with
Direct-Link and Two-Hop
schemes from a physical layer perspective is summarized in Table~\ref{tab:DF}.
\begin{table}
\caption{Summary of the phases for Direct-Link, Two-Hop and Decode-and-Forward
protocols from a physical layer perspective.}
\label{tab:DF}
\begin{center}
\begin{tabular}{c||c|c|c}
	&	Direct-Link	&	Two-Hop	&	Decode-and-Forward \\
\hline
Phase 1 &	\multicolumn{3}{c}{$\nnear\rightarrow\nap$}		\\ \hline
Phase 2	&	\multirow{2}{*}{$\nfar\rightarrow \nnear$} & $\nfar\rightarrow\nnear$ & $\nfar\rightarrow \nnear,\nap$\\
Phase 3 &						   &
$\nnear\rightarrow\nap$ &$%\textcolor{red}{\nfar},
\nnear\rightarrow \nap$ \\ \hline
\end{tabular}
\end{center}
\end{table}
We can define the achievable rate for this protocol as:
\begin{align}
\cdf=&\max_T\min\left\{
\stime_\snnear\log\lefto(1+\frac{\power}{\beta^\gamma\ttime_\nnear}\right),
\stime_\snfar\log\lefto(1+\frac{\power}{\betam^\gamma\ttime_\nfar}\right),
\right.
\nonumber\\
&\left.
\stime_\snfar\log\lefto(1+\frac{\power}{\ttime_\nfar}\right)+
\stime_\srelay\log\lefto(1+\frac{\power}{\beta^\gamma\ttime_\nnear}
\right)
\right\}.
\label{eq:decodeForwardFA}
\end{align}
The first term in~(\ref{eq:decodeForwardFA}) corresponds to $\nnear$ transmitting its
own packet to $\nap$ during $\stime_\snnear$.
The second and third terms correspond to the packet transmission of $\nfar$
using Decode-and-Forward protocol during $\stime_\snfar$ (Phase~2 in
Table~\ref{tab:DF}) and $\stime_\srelay$ (Phase~3).
There is a simple interpretation of this two-phase transmission. In $\stime_\snfar$,
the packet is completely transmitted to $\nnear$. This is guaranteed by
the second term in~(\ref{eq:decodeForwardFA}).
Then, the transmission $\nfar$--$\nap$ during
$\stime_\snfar$ and the transmission $\nnear$--$\nap$ during $\stime_\srelay$
can be interpreted as the transmission of data over two parallel AWGN channels
\cite{Cover2006}. The sum in the third term of (\ref{eq:decodeForwardFA}) then follows immediately
as the maximum mutual information between $(x_F,x_N)$ and $y_A$ from Eqs.~(\ref{eq:channelModelFirst})-(\ref{eq:channelModelLast}). Note that
the last two terms can be seen as a
special case of~\cite[Prop. 2]{Host-Madsen2005}, but differ from the SIMO
interpretation of the corresponding protocols II in~\cite{Nabar2004} and
Decode-and-Forward as defined in \cite{Laneman2004}.

From a MAC perspective, we adopt the same coordination strategy as in the
Two-Hop case, which is described in Remark~\ref{re:remark-queue}.
In each term in (\ref{eq:decodeForwardFA}), the
transmission time $\ttime$ can be strictly larger than $\stime$
because of protocol overhead such as acknowledgments (ACK) and packet headers or
because of collision when node $\nfar$ and node $\nnear$ are transmitting at the same time, which
can lead to interference between the transmissions that cannot be resolved by
the receiving node. Acknowledgement signals can resolve collisions such that
in each phase, the receiving node transmits an ACK if it can successfully decode the message.
After some timeout, if the source node did not receive ACK, the packet is
considered lost
and the source node retransmits the packet.
In our analysis of Section~\ref{sec:throughput}, collisions of ACK
transmissions are neglected. The reason for this assumption is that the
duration of ACK messages is very short compared to the transmission duration of
payload packets. 

\subsection{Decode-Idle-Forward}

As we shall see in Section~\ref{sec:throughput},
the naive (basic) Decode-and-Forward protocol suffers significantly
from the contention between nodes $\nfar$ and $\nnear$.
A simple but efficient strategy consists of using at node $\nfar$ the ACK signal sent by
$\nap$ to $\nnear$ right after Phase~3. Once $\nfar$ receives ACK from $\nnear$ after Phase~2,
$\nfar$ stays idle until receiving ACK from the access point $\nap$. Note that the protocol has
to ensure that $\nap$ sends ACK packet to $\nnear$ at a rate sufficiently low
such
that $\nfar$ can decode it. Once $\nfar$ gets ACK from $\nap$, $\nfar$ starts to compete again
for the channel access.

\subsection{Decode-Straightforward}

The Protocol~Decode-Idle-Forward can be further improved by noting that when
$\nfar$
is idle, $\nnear$ does not need
to compete for the channel access but can directly forward the message
(Phase~3). Clearly, this strategy
is only valid in the network model of Fig.~\ref{fig:terminals}. For larger
networks, $\nnear$ has still to compete for the channel access
with all other nodes except $\nfar$.

\section{Throughput Analysis}
\label{sec:throughput}
The purpose of this section is to calculate the max-min throughput
\eqref{eq:capacityDirectLast}, \eqref{eq:capacityTwohop}, and
\eqref{eq:decodeForwardFA} for the different MAC protocols that we proposed in
Section~\ref{sec:cooperative-protocols}. Maximization
over the time division parameters \eqref{eq:throughputParameter} cannot be
performed directly because of the interdependency between transmission times
$\ttime$ (which include collisions) and the successful transmission times
$\stime$ (which exclude collisions). We resolve this interdependency along the
lines of \cite{Bianchi2000}: First, we describe the network communication
system in terms of the independent parameters \textit{packetsize} and
\textit{transmission probability}. We then express the time division variables
\eqref{eq:throughputParameter} as functions of these parameters and maximize
\eqref{eq:capacityDirectLast}, \eqref{eq:capacityTwohop}, and
\eqref{eq:decodeForwardFA} over these parameters. In the low and high SNR
regimes, this maximization can be performed analytically by using asymptotic
approximations for \eqref{eq:capacityDirectLast}, \eqref{eq:capacityTwohop}, and
\eqref{eq:decodeForwardFA}; in the medium SNR range solutions can be found
numerically. We start by defining the aforementioned parameters.
\subsubsection{Packetsizes $\fpacket, \npacket,\rpacket$} The transmitters can
adjust the size of transmitted packets. For $\nfar$ and $\nnear$ transmitting
their own packets and $\nnear$ relaying, we denote the corresponding
packetsizes by $\fpacket$, $\npacket$, and $\rpacket$, respectively. We
arbitrary normalize the packetsizes such that $\fpacket+\npacket+\rpacket=1$
for sake of simplicity. As previously mentioned, DCF adopts an exponential
backoff scheme with a discrete time backoff scale. Since we normalize the
packet sizes, the corresponding timeslot duration has to be normalized
accordingly. We denote the normalized timeslot duration by $\sigma$. A typical
value would be $\sigma=50\mu\mathrm{s}/(3\cdot 8184\mu\mathrm{s})\approx 0.002$
\cite{Bianchi2000}, where the value $8184\mu\mathrm{s}$ reflects the average
packetsize for the three types of transmission.
\subsubsection{Probability of transmission $\tau$} Following
\cite{Bianchi2000}, the key modelling step is to assume that the network is in
steady state and that in any arbitrary phase, each node is transmitting with a
probability of $\tau$. For DCF, $\tau$ was calculated in \cite{Bianchi2000} in
terms of minimum contention window size,
number of backoff stages, and number of nodes competing for the channel. For
simplicity, we directly use $\tau$ as a protocol parameter over which throughput
is maximized. When both $\nfar$ and $\nnear$ are competing for the
channel, the probabilities of success, collision, and idle state can be 
calculated as
\begin{align}
\psucc=\tau(1-\tau),
\quad
\pcoll=\tau^2,
\quad
\pidle=(1-\tau)^2.\label{eq:transitionProbabilities}
\end{align}
A collision occurs when both $\nfar$ and $\nnear$ are transmitting at the same
time. Since both cannot send and receive simultaneously, they have
to finish their transmission before being able to detect collision.
Therefore, the duration of collision $\cpacket$ is given
by $\max\{\fpacket,\npacket\}$.
\subsection{Calculation of Throughput}
In the following, we express the time division
variables \eqref{eq:throughputParameter} as a function of packetsize and
transmission probability for Direct-Link and the three cooperative protocols
proposed in Section~\ref{sec:cooperative-protocols}. The three cooperative
protocols can readily be used for Two-Hop, with the only difference that for
Two-Hop, $\nap$ will discard what it receives from $\nfar$. Therefore, the
derived formulas for the time division variables \eqref{eq:throughputParameter}
can directly be used for the corresponding Two-Hop schemes.
\subsubsection{Random Access Direct-Link}
\begin{figure*}
\footnotesize
\centering
\psfrag{p11}{$p_{11}$}
\psfrag{p22}{$p_{22}$}
\psfrag{p12}{($p_{12}$)}
\psfrag{p21}{($p_{21}$)}
\psfrag{(a)}{(a)}
\psfrag{(b)}{(b)}
\includegraphics[width=0.8\textwidth]{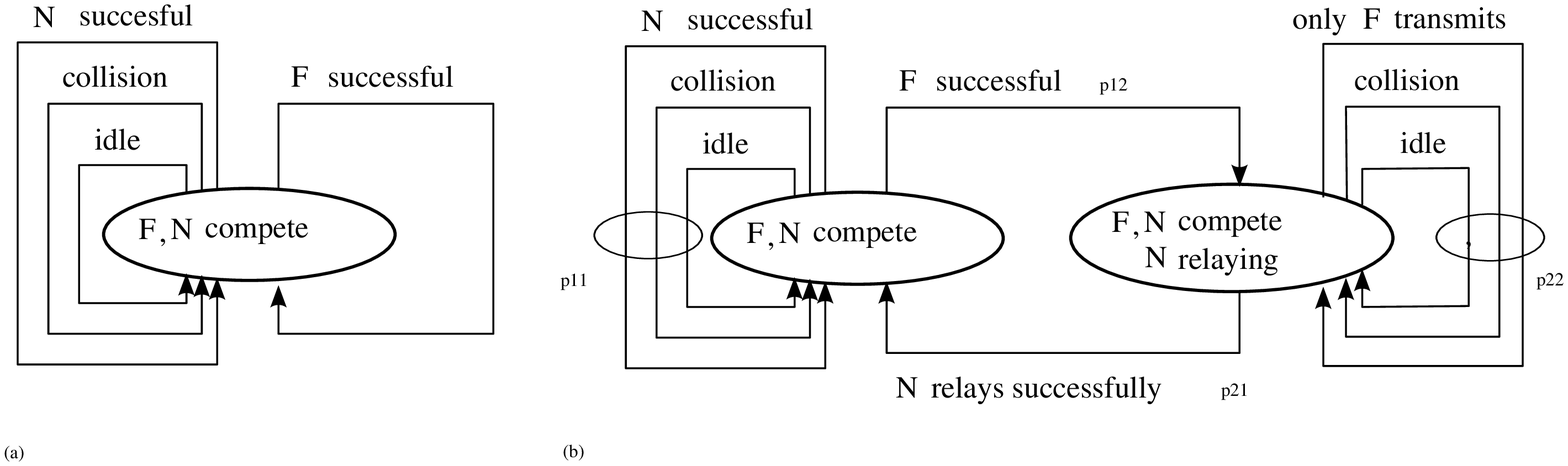}
\caption{Channel state diagram for Direct-Link (a) and Naive
Decode-and-Forward (b). When the transition phases ``only $\nfar$ transmits''
and ``collision'' are removed from state 2 in (b), the diagram illustrates
Decode-Idle-Forward. If in addition the transition phase ``idle'' is removed
from state 2, the resulting diagram illustrates Decode-Straightforward. Note
that the transition probabilities have to be adapted appropriately.}
\label{fig:states}
\end{figure*}
Both $\nfar$ and $\nnear$ are constantly competing for channel
access. As illustrated in Fig.~\ref{fig:states}(a), there are four different
transition phases:~successful transmission of $\nnear$, successful 
transmission of $\nfar$, collision, and idle mode. For Direct-Link,
$\nnear$ never relays, so $\rpacket=\stime_\srelay=0$ and
$\fpacket+\npacket=1$. By using the
probabilities from \eqref{eq:transitionProbabilities}, the expected duration $t$
of a transition phase is given by
\begin{align}
t&=\psucc\npacket+\psucc\fpacket+\pcoll\cpacket+\pidle\sigma\\
&=\tau(1-\tau)+\tau^2\max\{\fpacket,\npacket\}+(1-\tau)^2\sigma.
\label{eq:calcDirectLink1}
\end{align}
The average time $\nfar$ successfully transmits in a transition phase is
\begin{align}
\tsuccf=\tau(1-\tau)\fpacket.\label{eq:calcDirectLink2}
\end{align}
Using \eqref{eq:calcDirectLink1} and \eqref{eq:calcDirectLink2} in
\eqref{eq:throughputParameter}, the fraction of time $\stime_\snfar$ when
$\nfar$ is successfully transmitting can be expressed as
\begin{align}
\stime_\snfar=\frac{\tsuccf}{t}&=\frac{\tau(1-\tau)\fpacket}{
\tau(1-\tau)+\tau^2\max\{\fpacket,\npacket\}+(1-\tau)^2\sigma}
\label{eq:calcDirectLink3}
\end{align}
which is completely defined by the new set of parameters packetsize 
and transition probability as introduced at the beginning of this section.
Similarly 
\begin{align}
\ttime_\nfar=\frac{\tau\fpacket}{t},
\quad
\stime_\snnear=\frac{\tau(1-\tau)\npacket}{t},
\quad
\ttime_\nnear=\frac{\tau\npacket}{t}.\label{eq:calcDirectLink4}
\end{align}
We can use \eqref{eq:calcDirectLink3} and \eqref{eq:calcDirectLink4} to express
the time division variables in \eqref{eq:capacityDirectLast}. The
maximization problem over
$\{\ttime_\nfar,\ttime_\nnear,\stime_\snfar,\stime_\nnear\}$ has been turned
into a maximization problem over $\{\fpacket,\npacket,\tau\}$ subject
 to the constraints $\fpacket+\npacket=1$ and $0\leq\tau\leq 1$. It can
now easily be solved numerically.

Since the calculations are quite similar, we will only calculate
$\stime_\nfar$ for the remaining protocols. The other corresponding time
division variables can be expressed by packetsize and transmission
probability in an analogous way.
\subsubsection{Naive Decode-and-Forward Approach}
The network can be in the two states ``$\nfar,\nnear$ compete'' and
``$\nfar,\nnear$ compete, $\nnear$ relaying'', with which we associate the state
probabilities $\pi_1$ and $\pi_2$, respectively. See 
Fig.~\ref{fig:states}(b) for an illustration. The
transition probabilities between the two states are
\begin{align}
p_{12}&=p_{21}=\psucc,
\quad
p_{11}=p_{22}=1-\psucc
\end{align}
which implies $\pi_1=\pi_2=1/2$. The expected transition phase duration is
$t=\tsucc+\tcoll+\tidle$ with
\begin{align}
\tsucc&=\pi_1(\psucc\npacket+\psucc\fpacket)+\pi_2\psucc\rpacket\\
\tcoll&=\pi_1\pcoll\max\{\fpacket,\npacket\}+
\pi_2(\pcoll\max\{\fpacket,\rpacket\}+\psucc\fpacket)\\
\tidle&=\pi_1\pidle\sigma+\pi_2\pidle\sigma.
\end{align}
By using $\tsuccf=\pi_1\tau(1-\tau)\fpacket$, $\stime_\snfar$ in
\eqref{eq:throughputParameter} becomes
\begin{align}
\stime_\snfar&=\frac{1}{2}\tau(1-\tau)\fpacket\left[
\tau(1-\tau)+\frac{1}{2}\tau^2\max\{\fpacket,\npacket\}+\right.\nonumber\\
&\;\left.\frac{1}{2}\tau^2\max\{\fpacket,\rpacket\}+\frac{1}{2}
\tau(1-\tau)\fpacket+(1-\tau)^2\sigma\right]^{-1}
\label{eq:nThroughputCooperative}.
\end{align}
If we assume $\max\{\fpacket,\rpacket\}\approx\max\{\fpacket,\npacket\}$, the
main differences between $\stime_\snfar$ for Direct-Link
\eqref{eq:calcDirectLink3} and $\stime_\snfar$
for Naive Decode-and-Forward consists in the factor of $1/2$ in the
numerator and the term $\tau(1-\tau)\fpacket$ in the denominator, which
both result from the ``queue collision'' in state 2. It occurs when $\nfar$
successfully gains channel access, but $\nnear$ is ignoring the
transmission since it is still trying to forward the previous packet of
$\nfar$.
\begin{figure*}
\footnotesize
\centering
\subfloat[Throughput improvement for the three Decode-Forward protocols. 
Throughput $\cdf$ (\ref{eq:decodeForwardFA}) developed in
Section~\ref{sec:cooperative-protocols} is compared to the conventional approach
which consists of selecting the scheme Direct-Link or Two-Hop with highest
throughput (\ref{eq:C_nocoop}). The values are numerically calculated for
$\beta=0.5$, $\sigma=0.002$, and $\gamma=2$ following the procedure of
Section~\ref{sec:throughput}.]{
\includegraphics[width=0.47\textwidth]{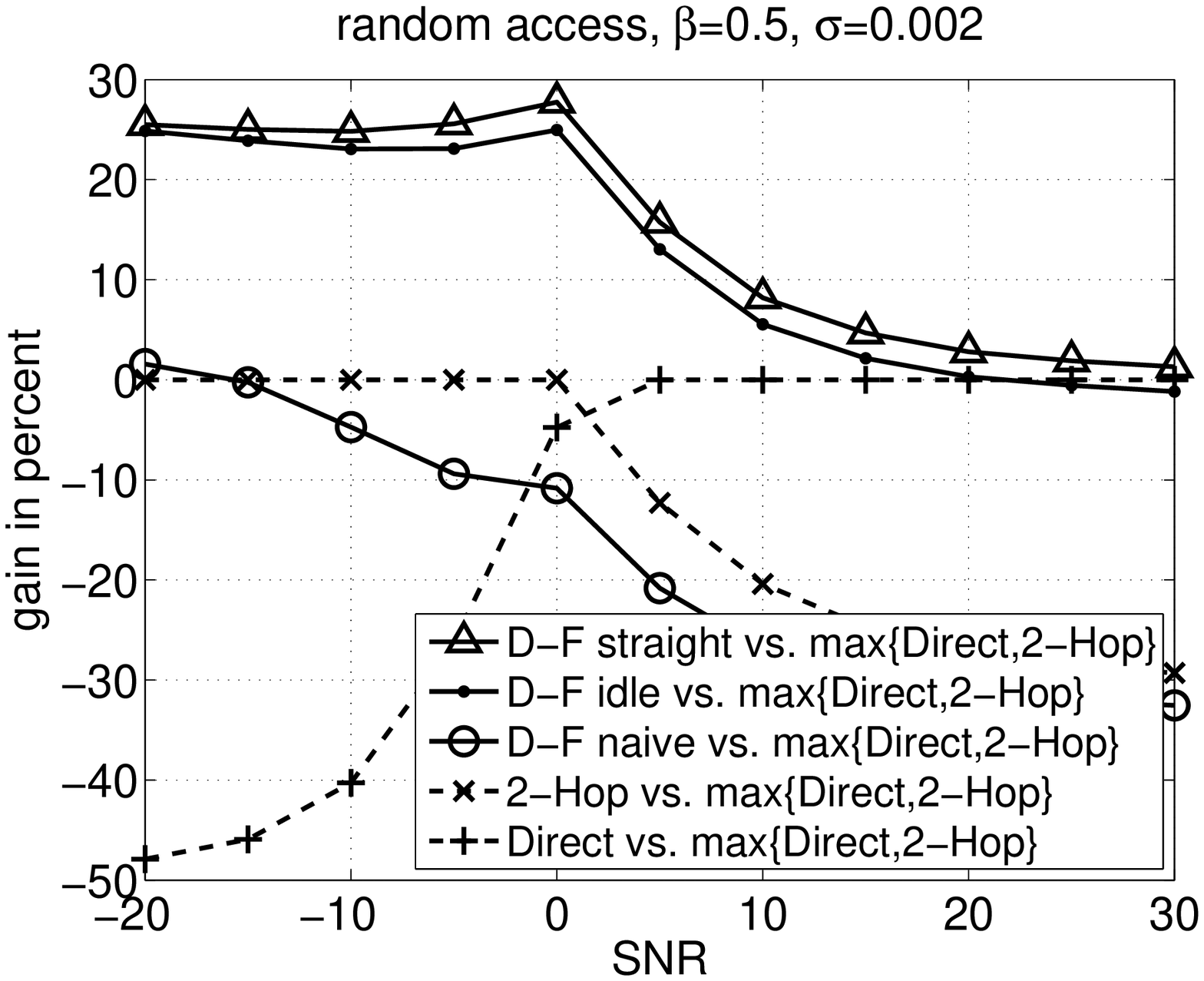}
\label{fig:ra05beta}
}
\hfill
\subfloat[Throughput improvement for the three Decode-and-Forward protocols.
Throughput $\cdf$ (\ref{eq:decodeForwardFA}) developed in
Section~\ref{sec:cooperative-protocols} is compared to
the conventional approach which consists of selecting the scheme Direct-Link
or Two-Hop with highest throughput (\ref{eq:C_nocoop}).
The values are numerically calculated for SNR$=0.5$~decibels, $\sigma=0.002$,
and $\gamma=2$ following the procedure of Section~\ref{sec:throughput}.]{
\includegraphics[width=0.47\textwidth]{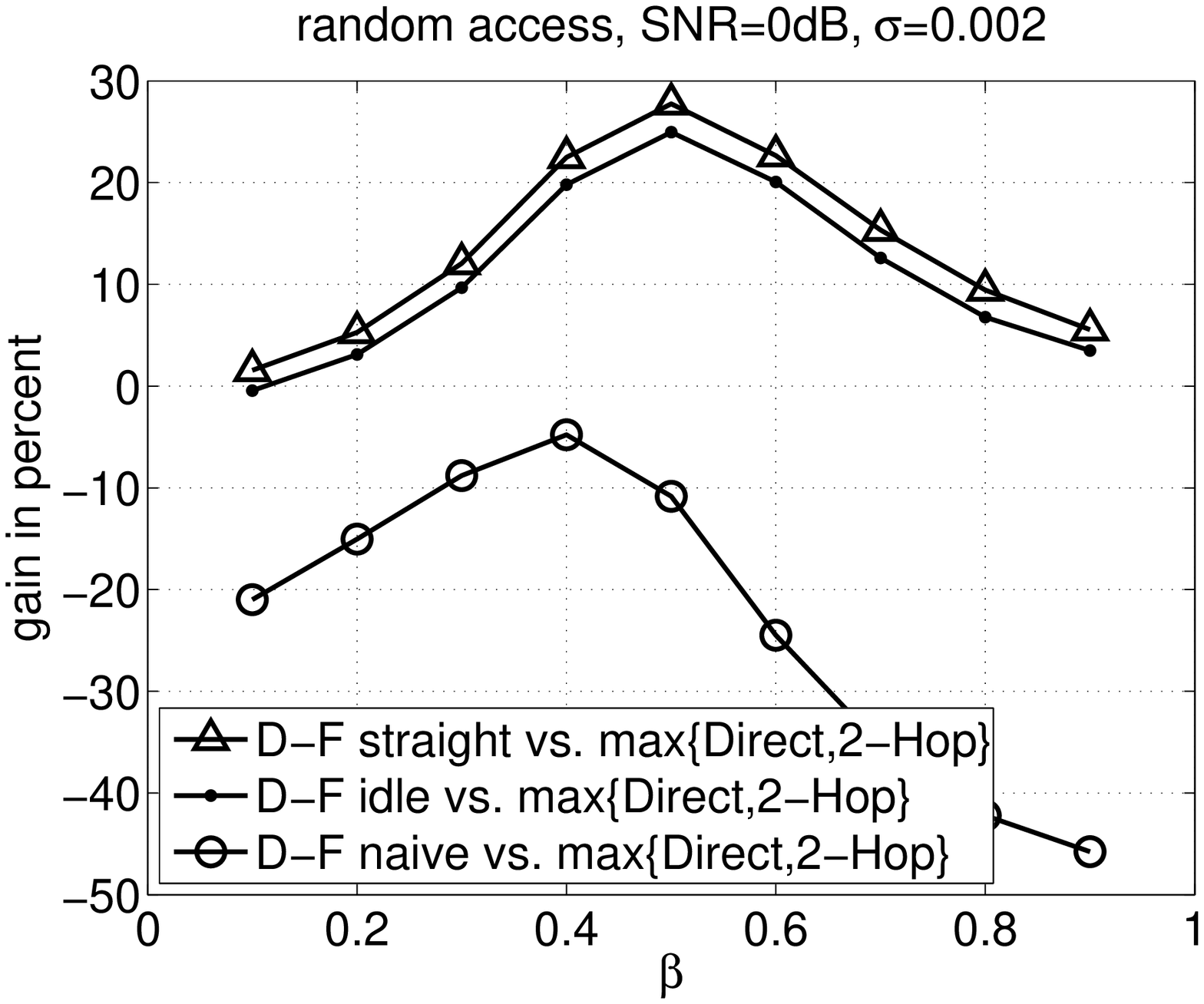}
\label{fig:ra0dB}
}
\end{figure*}

\subsubsection{Decode-Idle-Forward}
For Decode-Idle-Forward, the node $\nfar$ remains idle in state 2 in
Fig~\ref{fig:states}(b) and the state transition probabilities for state 2 are
given by $p_{21}=\tau$ and $p_{22}=1-\tau$. Consequently
 $(\pi_1,\pi_2)\propto(1,1-\tau)$ and the parameters for the expected phase
duration $t=\tsucc+\tcoll+\tidle$ are given by
\begin{align}
\tsucc&=\pi_1(\psucc\npacket+\psucc\fpacket)+\pi_2\tau\rpacket\\
\tcoll&=\pi_1\pcoll\max\{\fpacket,\npacket\}\\
\tidle&=\pi_1\pidle\sigma+\pi_2(1-\tau)\sigma.
\end{align}
Because $\tsuccf=\pi_1\tau(1-\tau)\fpacket$,
\begin{align}
\stime_\snfar=\frac{\tau(1-\tau)\fpacket}{\tau(1-\tau)+\tau^2
\max\{\fpacket,\npacket\}+2(1-\tau)^2\sigma}
\label{eq:nThroughputIdleForward}.
\end{align}
Compared to Direct-Link, there is an additional factor of two for the idle
timeslot $\sigma$ in the denominator.
\subsubsection{Decode-Straightforward}
For Decode-Straightforward, when $\nfar$ successfully transmits its packet to
$\nnear$, $\nnear$ knows that $\nfar$ will remain idle until $\nnear$ 
successfully forwards the packet to $\nap$. Therefore, it forwards the
packet directly with probability one to $\nap$. If we identify the effective
packet duration for $\nfar$ by $\fpacket+\rpacket$, the Decode-Straightforward
protocol is equivalent to the Direct-Link protocol from the MAC layer
perspective. Consequently, $\stime_\snfar$ is given by
\eqref{eq:calcDirectLink3} and the remaining time division variables in
\eqref{eq:throughputParameter} can easily be determined.

\section{Discussion}
\label{sec:discussion}
In this section, we illustrate the performance of the three cooperative
protocols Naive Decode-and-Forward, Decode-Idle-Forward, and
Decode-Straightforward developed in Section~\ref{sec:cooperative-protocols}
as a function of the distance $\beta$ between the relaying node $\nnear$ and
the access point $\nap$, and the average signal-to-noise ratio of the link
$\nfar$-$\nap$ defined in Section~\ref{sec:model-PHY}. The performance of these
protocols is compared to the conventional Direct-Link and Two-Hop schemes

Fig.~\ref{fig:ra05beta} shows the throughput improvement for the three
cooperative protocols compared to a conventional approach (\ref{eq:C_nocoop}),
which consists of selecting the scheme Direct-Link or Two-Hop with
highest throughput as in (\ref{eq:C_nocoop}). We favour Two-Hop in our
comparison by always using it with the MAC protocol of Decode-Straightforward,
which leads to least collisions. Node $\nnear$ is assumed to be
exactly in the middle of $\nfar$ and $\nap$ ($\beta=0.5$). We use the value of
$0.002$ for the normalized timeslot $\sigma$ as in Section~\ref{sec:throughput}.
The throughput improvement is shown for SNR ranging from $-20$ decibels to $30$ decibels.
Concerning the conventional approaches, Direct-Link scheme outperforms the
Two-Hop scheme in high SNR regime (SNR$>5$ decibels) whereas the Two-Hop scheme
outperforms the Direct-Link scheme in low SNR regime. Therefore, selecting the
conventional scheme with highest throughput is essential to be robust when
operating over a very large SNR range.
Note that for the Two-Hop scheme, we assume here that $\nfar$ remains idle as
long as $\nnear$ has to forward a packet from $\nfar$.
It is interesting that the Naive Decode-and-Forward protocol performs slightly
worse (approximatively $10$\%) than the Two-Hop scheme at any SNR (except for
very low values).
The degradation comes from the ``queue collision'' in state 2 in
Fig.~\ref{fig:states}(b) and cannot be compensated by exploiting the
information received by $\nap$ when $\nfar$ is transmitting. Queue collision
occurs at the MAC layer when $\nfar$ successfully gains channel access, but
$\nnear$ ignores the transmission since it is still trying to forward the
previous packet of $\nfar$.
For the cooperative protocols, the strategy that consists in maintaining $\nfar$
idle as long as $\nnear$ has to forward the missing information, provides
significant throughput gain at moderate and low SNR values (more than $20$\%).
In high SNR regime, the throughput gain versus the Direct-Link scheme becomes
less substantial. In this case, node $\nap$ receives most of the
information directly from $\nfar$ reducing the importance of the relay node.

Fig.~\ref{fig:ra0dB} shows the throughput improvement for the three cooperative
protocols compared to a conventional approach (\ref{eq:C_nocoop}) as a function
of the position of the intermediate node $\nnear$ in low SNR regime
(SNR~$=0$~decibel).
For the cooperative protocols that avoid the ``queue collision'' at Node
$\nnear$ (Decode-Idle-Forward and Decode-Straightforward), the throughput gain
over the conventional approach is maximal when $\nnear$ is located in the middle
between $\nfar$ and $\nap$. Interestingly, this throughput gain is equal to or
greater than $20$\% for $\beta$ ranging from $0.4$ to $0.6$. This is important
in larger networks where the selection of a relay is not trivial. For these
cooperative protocols, large throughput gains are observed even when the
selected relay is not precisely in the middle between $\nfar$ and $\nap$. As in
the previous setup, the degradation for the protocol Naive Decode-and-Forward
comes from the ``queue collisions'' and cannot be compensate by
exploiting the information received by $\nap$ when $\nfar$ is transmitting.

\section{Conclusions}
\label{sec:conclusions}
We proposed three cooperative protocols and compared them to the
conventional schemes Direct-Link and Two-Hop with respect to max-min
throughput. The key property of the proposed protocols is low complexity
achieved by random access. The first proposed protocol suffers from collision
and is outperformed by the conventional schemes. The second and third protocol
solve this problem in a distributed manner and outperform the conventional
schemes in the low SNR regime around $0$ decibel for a wide range of network
topologies. A natural application would be to increase the coverage of an access
point while maintaining the current max-min rate by using a cooperative
protocol.

We immediately note that our work has only scratched the surface in exploring
the issues in implementing cooperative systems. Natural next steps are to
investigate how the proposed protocols scale with an increasing number of nodes
in the network and what impact the relay selection problem has on the
achievable throughput.

%\appendix

%\input{appendix}

\bibliographystyle{IEEEtran}

\bibliography{IEEEabrv,Literatur}

\end{document}